\documentclass[10pt,conference]{IEEEtran}
\usepackage{cite}
\usepackage{amsmath,amssymb,amsfonts}
\usepackage{algorithm}
\usepackage{graphicx}
\usepackage{textcomp}
\usepackage{xcolor}
\usepackage[hyphens]{url}
\usepackage{fancyhdr}
\usepackage{hyperref}
\usepackage{balance}
\usepackage{listings}
\usepackage{caption}
\usepackage{comment}
\usepackage{xcolor}[table,xcdraw]
\usepackage{adjustbox}
\usepackage{setspace}
\usepackage{enumitem}
\usepackage{algpseudocode}
\usepackage{multicol}
\usepackage{tabularx}
\usepackage{array}
\usepackage{outlines}
\usepackage{soul}
\usepackage{flushend}
\usepackage{lipsum}
\usepackage{setspace}
\usepackage{enumitem}
\usepackage{float}
\usepackage{dblfloatfix}
\usepackage{multirow, makecell}
\makeatletter
\DeclareRobustCommand*\cal{\@fontswitch\relax\mathcal}
\makeatother

\newcommand{\hpcayear}{2024}

\renewcommand{\baselinestretch}{0.9933}

\title{Unlocking Hardware Security Assurance: The Potential of LLMs} 

\makeatletter
\newcommand{\linebreakand}{%
  \end{@IEEEauthorhalign}
  \hfill\mbox{}\par
  \mbox{}\hfill\begin{@IEEEauthorhalign}
}
\makeatother

\author{\IEEEauthorblockN{Xingyu Meng\IEEEauthorrefmark{1},
Amisha Srivastava\IEEEauthorrefmark{1}, Ayush Arunachalam\IEEEauthorrefmark{1}, Avik Ray\IEEEauthorrefmark{2},\\ 
Pedro Henrique Silva\IEEEauthorrefmark{3}, Rafail Psiakis\IEEEauthorrefmark{3}, Yiorgos Makris\IEEEauthorrefmark{1} and Kanad Basu\IEEEauthorrefmark{1}}

\IEEEauthorblockA{
\IEEEauthorrefmark{1}\textit{University of Texas at Dallas},
\IEEEauthorrefmark{2}\textit{Amazon Alexa},
\IEEEauthorrefmark{3}\textit{Technology Innovation Institute}}

\thanks {This research is supported by Technology Innovation Institute, Abu Dhabi, United Arab Emirates. ({\textit{Corresponding Author}:\textit{Xingyu Meng}}, Email: \textit{xxm150930@utdallas.edu}).
}
}

\fancypagestyle{camerareadyfirstpage}{%
  \fancyhead{}
  
  \fancyhead[C]{
    \ifdefined\aeopen
    \parbox[][12mm][t]{13.5cm}{\hpcayear{} IEEE International Symposium on High-Performance Computer Architecture (HPCA)}    
    \else
      \ifdefined\aereviewed
      \parbox[][12mm][t]{13.5cm}{\hpcayear{} IEEE International Symposium on High-Performance Computer Architecture (HPCA)}
      \else
      \ifdefined\aereproduced
      \parbox[][12mm][t]{13.5cm}{\hpcayear{} IEEE International Symposium on High-Performance Computer Architecture (HPCA)}
      \else
      \parbox[][0mm][t]{13.5cm}{\hpcayear{} IEEE International Symposium on High-Performance Computer Architecture (HPCA)}
    \fi 
    \fi 
    \fi 
    \ifdefined\aeopen 
      \includegraphics[width=12mm,height=12mm]{ae-badges/open-research-objects.pdf}
    \fi 
    \ifdefined\aereviewed
      \includegraphics[width=12mm,height=12mm]{ae-badges/research-objects-reviewed.pdf}
    \fi 
    \ifdefined\aereproduced
      \includegraphics[width=12mm,height=12mm]{ae-badges/results-reproduced.pdf}
    \fi
  }
  \fancyfoot[C]{}
}
\begin{document}
\maketitle

\ifdefined\hpcacameraready 
  \thispagestyle{camerareadyfirstpage}
  \pagestyle{empty}
\else
  \thispagestyle{plain}
  \pagestyle{plain}
\fi

\newcommand{\hpcaheight}{0mm}
\ifdefined\eaopen
\renewcommand{\hpcaheight}{12mm}
\fi


\begin{abstract}

System-on-Chips (SoCs) form the crux of modern computing systems. SoCs enable high-level integration through the utilization of multiple Intellectual Property (IP) cores. \textcolor{black}{However, the integration of multiple IP cores also presents unique challenges owing to their inherent vulnerabilities, thereby compromising the security of the entire system.} \textcolor{black}{Hence, it is imperative to perform hardware security validation to address these concerns.} The efficiency of this validation procedure is contingent on the quality of the SoC security properties provided. \textcolor{black}{However, generating security properties with traditional approaches often requires expert intervention and is limited to a few IPs, thereby resulting in a time-consuming and non-robust process.} To address this issue, we, for the first time, propose a novel and automated Natural Language Processing (NLP)-based Security Property Generator (NSPG). \textcolor{black}{Specifically, our approach utilizes hardware documentation in order to propose the first hardware security-specific language model, HS-BERT, for extracting security properties dedicated to hardware design.}
 \textcolor{black}{To evaluate our proposed technique, we trained the HS-BERT model using sentences from RISC-V, OpenRISC, MIPS, OpenSPARC, and OpenTitan SoC documentation.} When assessed on five untrained OpenTitan hardware IP documents, NSPG was able to \textcolor{black}{extract 326 security properties from 1723 sentences}. This, in turn, aided in identifying eight security bugs in the OpenTitan SoC design presented in the hardware hacking competition, Hack@DAC 2022.

\end{abstract}

\section{Introduction}
\label{intro}
Modern computing systems are built on System-on-Chips (SoCs), as they offer a high level of integration through the use of multiple Intellectual Property (IP) cores~\cite{farzana2019soc}. However, this also presents new security challenges, since vulnerabilities in one IP core may affect the security of the entire system~\cite{dessouky2019hardfails}. While software and firmware patches can address many hardware security vulnerabilities, some cannot be fixed and require extensive security assurance during the design process. Hence, hardware security validation is imperative to ensure the security and trustworthiness of the design. MITRE's Common Weakness Enumeration (CWE) for hardware categorizes commonly encountered security weaknesses in hardware designs, including issues with security flow, privilege and access control, reset control, memory and storage, peripherals and on-chip fabric, and debugging and test~\cite{CWECWE1113:online}.  
Commercial verification tools, such as JasperGold Security Path Verification, Mentor Questa Secure Check, and Tortuga Logic Radix, have been proposed for SoC security verification~\cite{JasperGo88:online,QuestaSe84:online,TortugaL60:online}. However, the effectiveness of these tools depends on the quality of the security properties. Therefore, these properties are critical components that provide resources for detecting vulnerabilities during SoC development.

\begin{figure}[tb!]
\centering
  \includegraphics[width=0.80\linewidth]{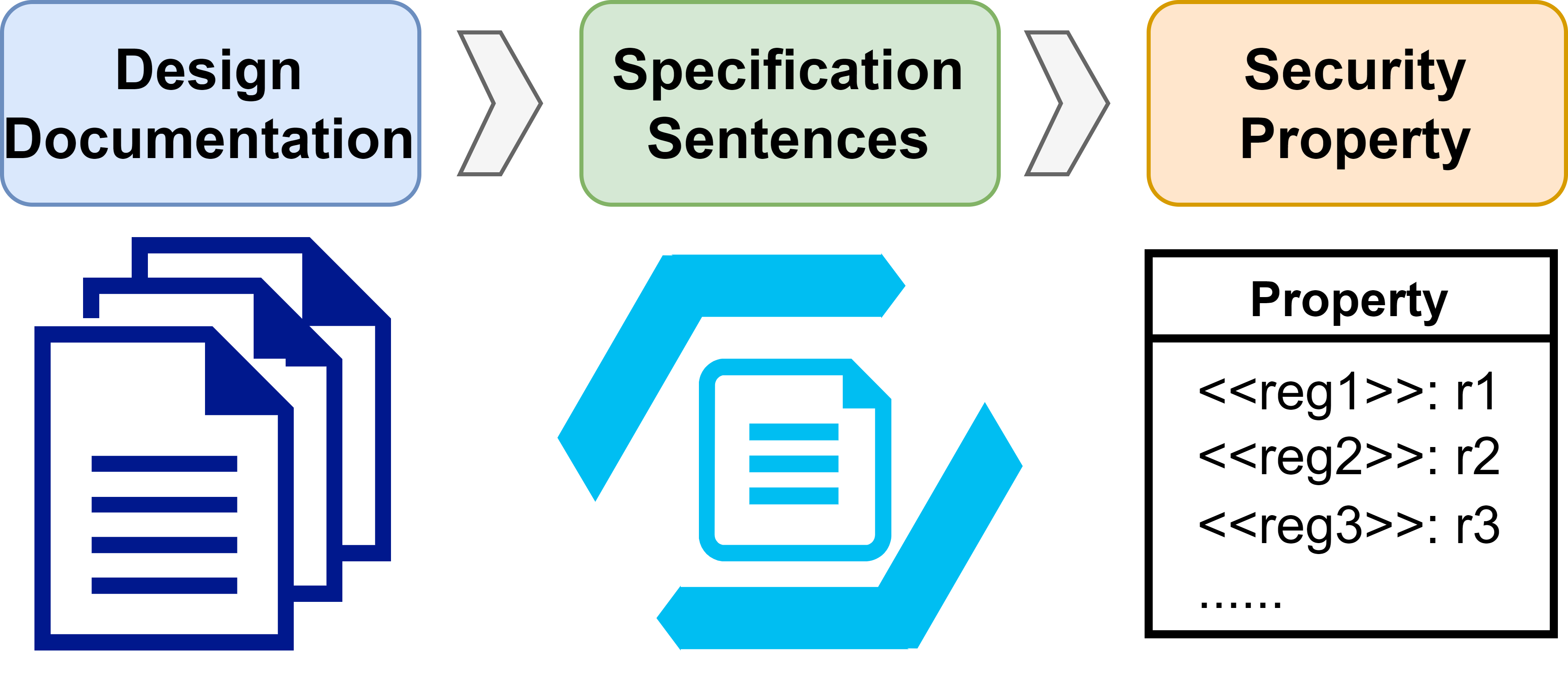}
  \caption{Security property generation from documents.}
  \label{fig:NLPplan}
  \vspace{-0.2in}
\end{figure}

Generating appropriate security properties for each specific design can be a challenge. Traditional approaches, which often require the experience of developers, are time-intensive and non-robust~\cite{dessouky2019hardfails}. Hence, a technique that is capable of systematically generating security properties for SoCs is needed to address this issue. On the other hand, major organizations such as RISC-V, MSP, and Arduino usually provide documentation describing processor functionalities and operation behaviors with explicit details~\cite{OpenTita77:online, RISCVIns68:online}. Therefore, as shown in Figure~\ref{fig:NLPplan}, we reason that it is possible to generate numerous security properties by analyzing operation details from these documents, which can be converted into security constraints at the Register-Transfer Level (RTL). We aim to achieve this by creating a language-based machine learning framework to extract essential information from the documentation. 

\begin{figure}[tb!]
\centering
  \includegraphics[width=0.85\linewidth]{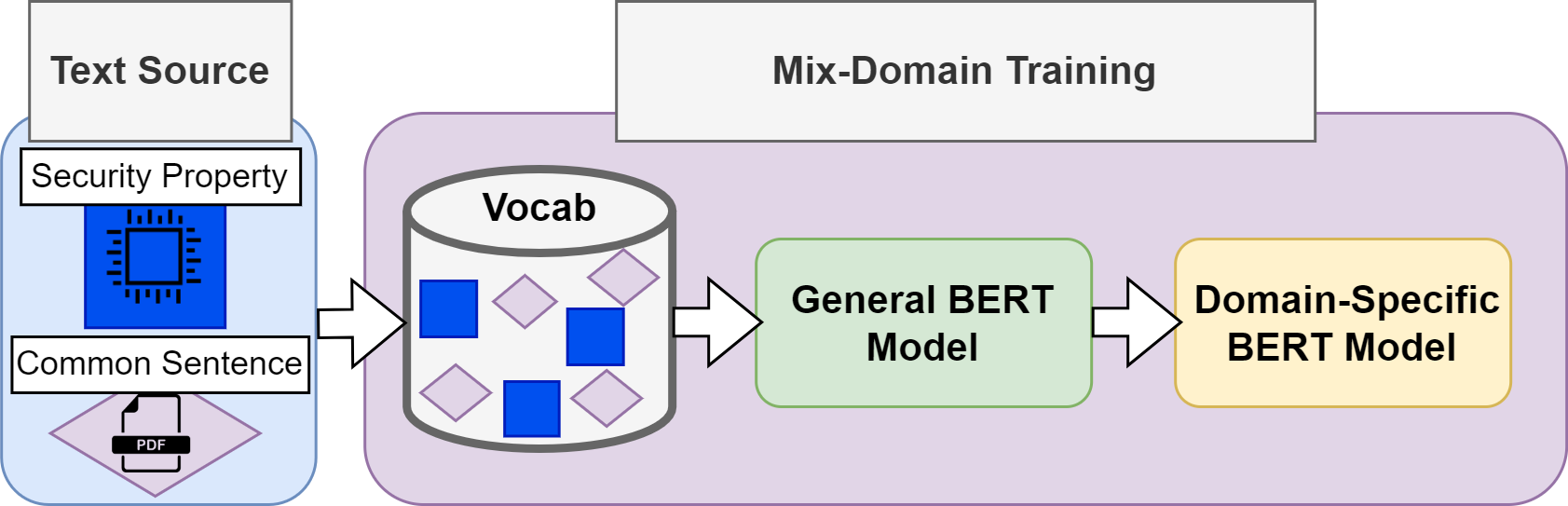}
  \caption{ Domain-specific BERT model.}
  \label{fig:NLP}
  \vspace{-0.1in}
\end{figure}

Existing research in the biomedical field, such as BioBERT, has shown that text and documents can provide essential information in specialized domains to fine-tune the general Bidirectional Encoder Representations from Transformers (BERT) language model for text generation and phrase classification~\cite{beltagy2019scibert,lee2020biobert}. Similarly, security property-related sentences in hardware documentation also furnish domain-specific terminology. Hence, we intend to apply this concept to the hardware security domain and develop an automatic security property extraction framework built on the BERT model, which is fine-tuned with domain-specific data, as shown in Figure~\ref{fig:NLP}. Our aim is to utilize this framework to extract each sentence in the documentation that can be potentially converted into security properties.
To this end, we developed a fully automatic hardware security property generator called NLP-based Security Property Generator (\textbf{NSPG}). 
NSPG applies data augmentation and masked language model to enhance the dataset and improve the learning process.
\textcolor{black}{We compared the output of NSPG with the expected information flow policies of these designs and found that the generated specifications were accurate and comprehensive, and if followed, would protect the designs from known and potential future attacks.} 
To the best of our knowledge, NSPG is the first property generation technique utilizing the NLP model and SoC documentation to generate security properties for hardware designs. Our contributions are summarized as follows:
\begin{enumerate}

   \item  \noindent\textcolor{black}{We propose an NLP-based security property generation framework, NSPG, to automatically mine security property-related sentences from the SoC documents, assisting in the detection of security vulnerabilities in RTL.}

   \item \noindent \textcolor{black}{We present a complete end-to-end framework with hardware domain-specific knowledge and data modification techniques to improve the performance of the proposed HS-BERT model by analyzing hardware documentation.} 

    
   \item \noindent NSPG is evaluated on five unseen OpenTitan design documents and all generated security properties are validated. Furthermore, we apply these properties to search for security violations in the OpenTitan design used in Hack@DAC 2022 and identify eight bugs~\cite{HACKDAC211:online}.


   \item \noindent When compared with ChatGPT, NSPG shows 15\% improvement on identifying security properties in OpenTitan SoC documentation~\cite{Introduc18:online}.
\end{enumerate}

The rest of this paper is organized as follows. Section~\ref{back} introduces the background of hardware security, and techniques in the NLP domain. Section~\ref{method} includes the proposed technique. Section~\ref{threat} introduces the SoC architecture, threat model, and security objectives used in this work.
Section~\ref{experiment} demonstrates the evaluation of the proposed technique. Section~\ref{discussion} discusses the capabilities of the proposed technique. Section~\ref{related} presents the related works on security property generation. Finally, Section~\ref{conculsion} concludes our paper.

\section{Background}
\label{back}

In this section, we will provide some background on hardware security verification, NLP, and the SoC designs used for developing the proposed NSPG framework.

\subsection{Hardware Security}

\subsubsection{Hardware Security Verification}

A plethora of techniques have been developed to ensure the security of software applications, either through source code or binary operations~\cite{dessouky2019hardfails}. 
However, the availability of commercial Electronic Design Automation (EDA) tools, which are specifically crafted for hardware security, is rare.
Hence, recently, researchers have focused on developing tools and methodologies to ensure hardware security. 
Nevertheless, all existing security verification techniques need robust security properties to validate the trustworthiness and robustness of the RTL~\cite{dessouky2019hardfails}. More details on hardware security verification approaches will be presented in Section~\ref{sec:sec_app}.

\begin{figure*}[t!]
\centering
  \includegraphics[width=0.9\textwidth]{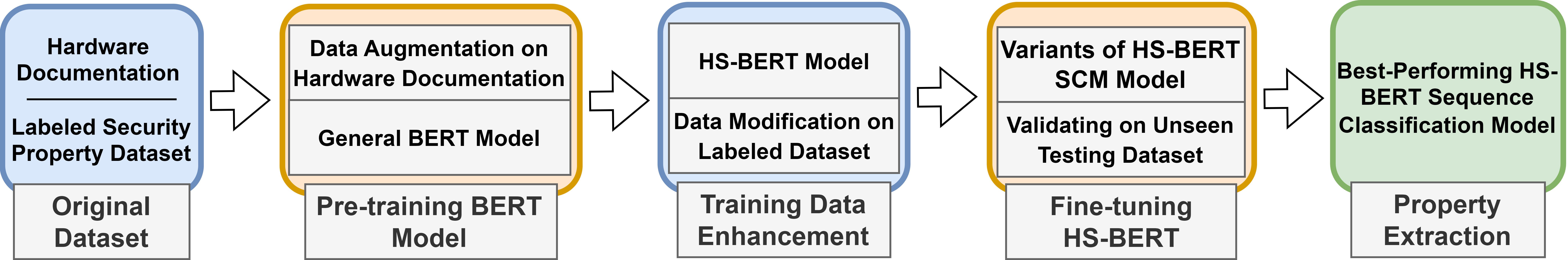}
  \caption{Proposed NSPG framework.}
  \label{fig:bertflow}
  \vspace{-0.2in}
\end{figure*}
\subsubsection{Hardware Design Documentation}

\textcolor{black}{In order to enhance the comprehensibility of the SoCs, it is imperative to have access to thorough SoC documentation, considering their complexity and dependence on reusable components}. {However, documentation is generally considered to provide limited value for any scientific analysis, and, thus overlooked by researchers.} 
\textcolor{black}{In reality, technical documentation is an important part of the overall product and should be prioritized along with the design and implementation stages. In more than one instance, insufficient documentation has been regarded as the major reason for design failure~\cite{bentley2001validating}.}
Additionally, various tools have been introduced to generate system documentation more efficiently. For instance, ``Javadoc'' and ``CppDoc'' 
generate API documentation in HTML from the comments in the source code~\cite{JavadocT88:online, GitHubsh47:online}.
\subsubsection{Hardware Security Property}
\label{sec:hard_sec}
\textcolor{black}{The specification of security properties usually varies from the method of security analysis, producing specifications that are unique to the selected verification tools or models~\cite{mana2008towards}.} 
\begin{center}
\begin{minipage}{0.95\linewidth}
\vspace{-0.1in}
\begin{lstlisting}[caption={Security Property and SystemVerilog Assertion},label={lst:1}, escapechar = @]
Security Property Description:
If the AES unit wants to finish encryption/decryption
of a data block but the previous output data has not
yet been read by the processor, AES unit is stalled.

SystemVerilog Assertion:
assert  property (                                  
  @\@@(posedge clk) disable iff (rst)// Security Property
    aes.done |-> aes.out @==@ @\$@past(aes.key)
  )
  else                           // Error Message
    @\$@error("@\%@m previous key has not been read");
\end{lstlisting}
\end{minipage}
\end{center}
\textcolor{black}{Listing~\ref{lst:1} presents an example of SystemVerilog assertion based on the description of the security property. It shows that in order to generate an appropriate term for a verification mechanism, the descriptions of security properties must contain strong reasoning and details of the operation.
Listing~\ref{lst:2} shows a paragraph from the OpenTitan AES document~\cite{Introduc18:online}. The sentences marked in blue are security related.}

\begin{center}
\begin{minipage}{0.95\linewidth}
\vspace{-0.1in}
\begin{lstlisting}[caption={A example paragraph in AES Document},label={lst:2}, escapechar = @]
Also, there is a back-pressure mechanism for the
output data.@\aftergroup\bluecolor@ If the AES unit wants to finish the 
encryption/decryption of a data block but the previous
output data has not yet been read by the processor, 
the AES unit is stalled. @\aftergroup\blackcolor@It hangs and does not drop 
data. The order in which the output registers are read 
does not matter. @\aftergroup\bluecolor@Every output register must be read
at least once for the AES unit to continue.@\aftergroup\blackcolor@ This is 
 the default behavior. @\aftergroup\bluecolor@It can be disabled by setting 
the MANUAL_OPERATION bit in CTRL_SHADOWED to 1.@\aftergroup\blackcolor@
\end{lstlisting}
\end{minipage}
\end{center}

\textcolor{black}{It is obvious that the security properties have more distinguished contexts, such as the usage of relation conjunctions and domain-specific terms, compared to other sentences in the paragraph. Therefore, it is possible to distinguish the security property-related sentences, which present precise definitions of the schematic, keywords of the processes, or relations between several entities.
Hence, we can emphasize these aspects when fine-tuning the language model by enhancing and identifying the security property-related context in each sentence. Further details about the security property extraction techniques utilized in our framework will be presented in Section~\ref{sec:method}.}

\subsection{Natural Language Processing}

\textcolor{black}{NLP is a field of study that
focuses on applying computational techniques to understand, learn, and generate human language content.} 
\textcolor{black}{NLP plays a crucial role in various industries and has a wide range of applications, ranging from real-time translation and social media search engines to sentiment analysis~\cite{feng2021survey}.} 
In this section, we will discuss the NLP facets that are integrated into NSPG.


\subsubsection{BERT Model}

Most state-of-the-art natural language models are built on transformer architectures, such as  Bidirectional Encoder Representations from Transformers (BERT), which are effective at modeling long-range dependencies in text~\cite{gillioz2020overview}. These models utilize a multi-layer, multi-head self-attention mechanism, and contextual embeddings which allow for efficient parallel computation on GPUs~\cite{devlin2018bert}. \textbf{We utilize the BERT masked language model to comprehend the hardware documentation and BERT sequence classification model to extract the security property-related sentences.}

\subsubsection{Data Augmentation}

Data augmentation (DA) is a method of enhancing the diversity of training data without collecting more data. It involves adding modified copies of existing data or creating synthetic data to act as a regularizer and reducing overfitting during the training of machine learning models~\cite{feng2021survey}. 
\textbf{Since this is the first work that utilizes design documentation for hardware security, our data samples are limited to open-source documentation. Thus, we will apply DA approaches, as discussed in Section~\ref{method}.}

\subsubsection{Masked Language Modeling}

Masked Language Modeling (MLM) is a self-supervised learning method used in state-of-the-art BERT models, such as SciBERT or RoBERTa~\cite{beltagy2019scibert, devlin2018bert}. 
MLM is used to improve the ability to comprehend the context and relationships between each word in a sequence, such as the security property demonstrated in Listing~\ref{lst:1}. 
\textbf{We will utilize MLM to modify the sequences from the document and provide additional data for fine-tuning.}

\subsubsection{Sequence Classification Modeling}
\label{sec:SCM}
Sequence Classification Modeling (SCM) divides sequences into predefined sentiment categories~\cite{devlin2018bert}. 
\textbf{Our proposed framework, NSPG, will apply this model to categorize each sequence of sentences and identify if they belong to security property or non-property descriptions in the documents.}



\section{Can Leveraging Large Language Models Foster Hardware Security Assurance?}
\label{method}
\begin{scriptsize}
\begin{table*}[!b]
\renewcommand{\arraystretch}{1.4}
\caption{Data Augmentation for sentence in documents. The first row shows the original sentence in the document, the rest shows an example sentence after the random swap, random deletion, synonym replacement, and random insertion.}
\label{tab:DA}
\begin{adjustbox}{width=0.95\textwidth,center}
\begin{tabular}{ll}
\hline \hline
                          & \multicolumn{1}{l}{\textbf{Sentence}}                                                                                                 \\ \hline \hline
\textbf{Original}         & If some hang condition were to occur when in this mode, the main state machine debug register should be read.                         \\
\textbf{Original with RS} & If some hang condition were to occur, the main state machine debug register should be read \underline{when in this mode}.             \\
\textbf{Original with RD} & If \st{[some]} hang condition were to occur when in this mode, the main state machine debug register should be read.                  \\
\textbf{Original with SR} & If some hang condition were to \underline{happen} when in this mode, the main state machine debug register should be read.            \\
\textbf{Original with RI} & If some hang condition were to occur when in this mode, the main state machine debug register should be read \underline{[immediately]}. \\ \hline \hline
\end{tabular}
\end{adjustbox}
\vspace{-0.1in}
\end{table*}
\end{scriptsize}

\label{sec:method}
In this section, we will demonstrate the details of each process flow in the proposed NSPG framework, as shown in
Figure~\ref{fig:bertflow}. The first step involves comprehending the context of the hardware domain from the hardware design documentation of OpenTitan and RISC-V, and generating the hardware security-specific BERT model. Next, we alter them using data augmentation and hardware domain-specific modification to create an enhanced dataset for fine-tuning the pre-trained BERT model. We compare the performance of various modified datasets used to fine-tune the classification model and select the best one for security property extraction.
The details of the data augmentation and pre-training hardware-domain BERT will be discussed in detail in Section~\ref{sec:datamod}.
The domain-specific data modification process and security property classification will be described in detail in Section~\ref{sec:spg}.

\subsection{Hardware Documentation Dataset}
\label{sec:dataset}
\textcolor{black}{
The sentences used for training and fine-tuning are extracted from various documentation and paragraphs similar to the example shown in Listing~\ref{lst:2}. We create three datasets, as follows:
(1) \textcolor{black}{15583 sentences from OpenTitan, RISC-V, OpenRISC, MIPS, and OpenSPARC documentation are used for pre-training the BERT model with MLM.} 
This dataset will be called ${\cal{D}}_{pre}$. 
(2) \textcolor{black}{To fine-tune the classification model, we manually labeled training samples consisting of 4427 sentences (out of which 2191 are security property-related sentences and the rest are non-property-related sentences) from the top module and test cases documentation in OpenTitan design, MIPS, OpenSPARC as well as non-privileged documentation for RISC-V.}  This dataset will be called ${\cal{D}}_{cls}$. 
(3) In order to simulate the scenario of processing unseen documentation, we also manually labeled additional 708 security property-related and non-property-related sentences to validate the fine-tuned SCM model. \textcolor{black}{It consists of 470 security properties and non-properties from OpenTitan AES and ADC documentation, 78 sentences from OpenRISC, and 160 sentences from RISC-V trace documentation.} All of these sentences are not included in ${\cal{D}}_{pre}$, and this dataset will be called ${\cal{D}}_{val}$.}

\subsection{Comprehending the Hardware Domain}
\label{sec:datamod}

\begin{figure}[tb!]
\centering
  \includegraphics[width=0.95\linewidth]{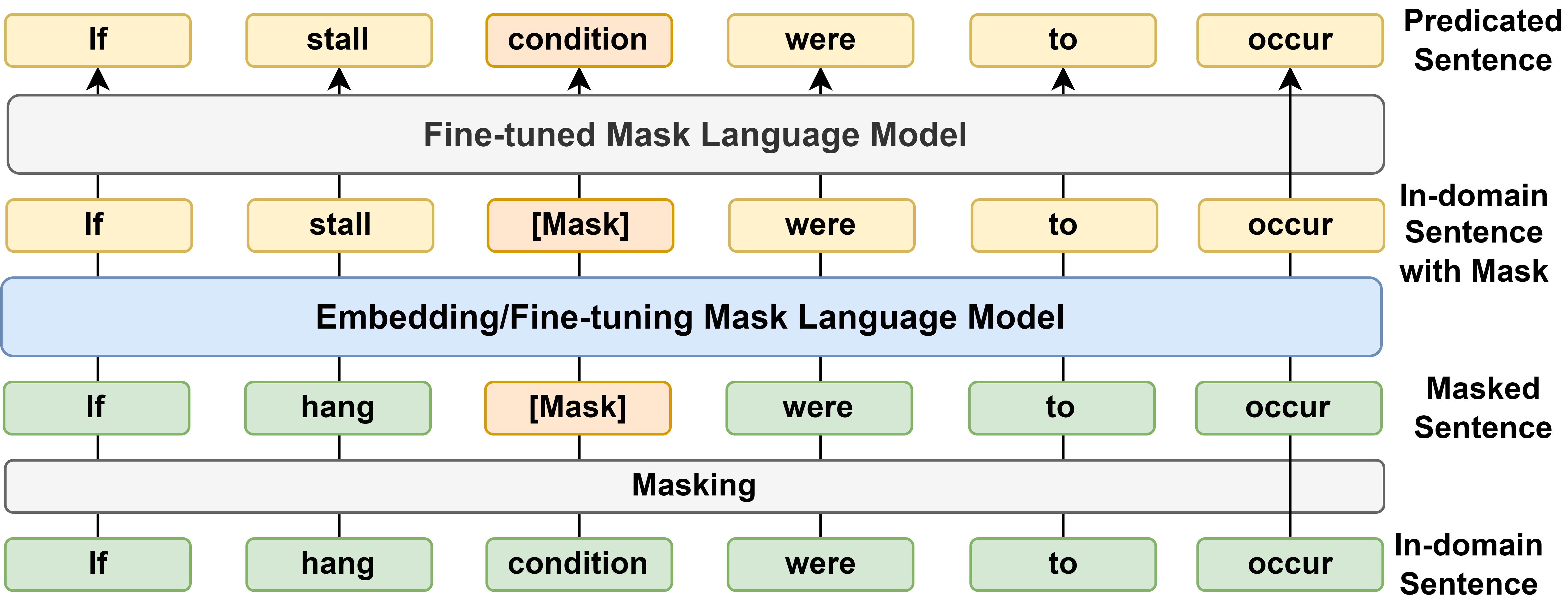}
  \caption{Masked language model.}
  \label{fig:BertModel}
  \vspace{-0.1in}
\end{figure}

First, the language model needs to recognize the context differences between security properties and regular sentences. General BERT model has been applied in various industries and has a wide range of applications, ranging from real-time translation and social media search engines to sentiment analysis~\cite{feng2021survey}.
Training the BERT model with its corresponding MLM could augment the performance of the language model by introducing contextual embeddings of the specific domain. \textcolor{black}{As shown in Figure~\ref{fig:BertModel}, during training, a certain percentage of tokens are randomly selected and replaced with a special token ([MASK] in Masked sentence). After training, the fine-tuned MLM will predict the original tokens based on the context provided by the remaining unmasked tokens, thus completing the in-domain sentence with appropriate phrases.} \textcolor{black}{The primary objective is to minimize the cross-entropy loss between the predicted tokens and their original counterparts.} For instance, in SciBERT and RoBERTa, by default, 15\% of the input tokens are chosen for masking, with 80\% probability of being replaced by [MASK], 10\% left unchanged and 10\% randomly replaced by another token from the vocabulary~\cite{beltagy2019scibert,devlin2018bert}.
Therefore, pre-training BERT model with hardware domain documentation will help it learn to represent the in-domain words based on the context of the other words in the sentence. However, since we have limited data samples for hardware domain-specific documentation (compared to 1.14M papers from Semantic Scholar to train Sci-Bert), data augmentation is needed to improve the dataset\cite{beltagy2019scibert}.

\subsubsection{Data Augmentation for Hardware Documentation}

\label{sec:da}
Many DA techniques have been proposed for NLP, including rule-based manipulations and generative approaches. Ideal DA techniques should be implemented seamlessly in order to further improve model performance. \textcolor{black}{Techniques that use trained models are more expensive to implement. However, they introduce more variance in data, thereby resulting in enhanced model performance. Although model-based techniques are designed to boost performance in downstream tasks, they are difficult to develop and use. This may result in overfitting or performance degradation when trained on out-of-domain examples.} 
DA is a key component for enhancing the quantity of in-domain data samples, which involves generating two correlated views of a data point in order to increase the amount and diversity of training data. Although it is important to ensure the diversity of the generated data, the structure, and synonym of the original sentence should not be modified. 

Most of the security properties involve the behaviors of two or more entities in the design. Therefore, the context of operational relations and hardware terminology needs to be preserved in fine-tuning MLM. We need to avoid the essential information of the operating entities and only apply DA to swap or replace the rest of the tokens in the sentence. We analyze four DA techniques: random swap, random deletion, synonym replacement, and random insertion. These are the most common data augmentation technique available for NLP model~\cite{feng2021survey}.  The closest sentiment texts are provided by Wordnet, a large lexical database of English, where synonyms are interconnected by means of conceptual-semantic, to generate the augmented data~\cite{miller1995wordnet}. 
The tokens in the texts are modified in the following ways to generate the augmented data. Table~\ref{tab:DA} presents examples of the techniques applied to a sentence from the DA task, as follows:

 \noindent{\textbf{Random Swap (RS)}}: \ul{We exchange the positions of two randomly selected phrases such as nouns and conjunctions, while maintaining the operation behavior and the content of the original sentence.} For example, as evident from Table~\ref{tab:DA}, RS takes parts of the sentence conjunction ``when in this mode'' and places them into a random spot, which is either before or at end of a conjunction in the original sentence. In case the sentence does not have multiple conjunctions, no augmented sentence will be generated.
 
 \noindent{\textbf{Random Deletion (RD)}}: \ul{To keep the essential context, we only delete one of the adjectives, determiners, or adverbs available in the sentence, which does not impact the operation descriptions.} For example, RD removes the definition term ``some'' from the sentence in Table~\ref{tab:DA}. When these terms are not found in the sentence, RD will not be applied.
 
 \noindent{\textbf{Synonym Replacement (SR)}}: \ul{In order to preserve the hardware components described in the sentence, we only replace verbs in the sentence with their closest synonyms obtained from the WordNet database.} It utilizes the database to find the closest semantic words for the sentence and replace the verb with the new ones. For example, in Table~\ref{tab:DA}, ``occur'' is replaced with its synonym ``happen'', and the sentence still presents the same operation. When the sentence only contains verbs such as ``is'' or ``are'', no augmented sentence will be generated.
 
 \noindent{\textbf{Random Insertion (RI)}}: For this operation, we first summarize all the adverbs that are used in the documentation, and \textcolor{black}{randomly select one to insert into the sentence near verbs}. \ul{To avoid altering the context, we randomly choose the most common adverb used in the documents.} For instance, insert the adverb 'immediately' before or after a randomly selected verb in the sentence, as shown in Table~\ref{tab:DA}.

 Each method tends to introduce diversity in the original sentence and increase possible word usage, thereby reducing data overfitting and boosting the generalization ability of the model. 
 \textcolor{black}{We train four BERT models using different augmented data (in combination with the original data) and evaluate their performance to determine the best model.} 

\subsubsection{Pre-training BERT with MLM}
\label{sec:mlm}

\begin{scriptsize}
\begin{table}[t!]
\caption{The performances of HS-BERT models with the original sentence from ${\cal{D}}_{pre}$ and after the random swap, random deletion, synonym replacement, and random insertion.}
\label{tab:mlm}
\begin{adjustbox}{width=1\columnwidth,center}
\begin{tabular}{cccc}
\hline
\hline 
     \textbf{Dataset}     &    \textbf{Training Samples}      & \textbf{Runtime} & \textbf{Perplexity} \\ \hline \hline
\textbf{Original Documents}  &  12472       & 57 mins          & 6.018                         \\ \hline
\textbf{Original \& RS}    &   24065    & 109 mins          & 5.018                          \\ \hline
\textbf{Original \& RD}    &   24946    & 114 mins          & 5.046                             \\ \hline
\textbf{Original \& SR}   &    24208    & 110 mins          & 5.029                          \\ \hline
\textbf{Original \& RI}   &    24931     & 113 mins          & 4.795                             \\ \hline
\end{tabular}
\end{adjustbox}
\end{table}
\end{scriptsize}

The purpose of MLM is to understand the detail of each sentence demonstrated in the document and predict the masked content. \ul{Hence, we will select the model having the lowest \textit{perplexity}, where a lower score indicates that the model has a better comprehension of the hardware domain and prediction of the masked word in the in-domain sentences}~\cite{moore2010intelligent}. We split the samples from ${\cal{D}}_{pre}$ into training and validation datasets with a ratio of 80\% and 20\%.
Each model is trained with samples from the OpenTitan, RISC-V, OpenRISC, MIPS, and OpenSPARC documentation, which consist of 12473 sentences and additional data from the DA task.
Table~\ref{tab:mlm} shows the runtime, and perplexity for each DA approach, as described in Section~\ref{sec:da}. \textbf{Note that the runtime differences are caused by different numbers of augmented data samples since the operation details of original sentences need to be preserved. We evaluate each pre-trained BERT model with the validation dataset consisting of 3112 sentences. As shown in Table~\ref{tab:mlm}, the DA task significantly improves the performance of the BERT model comprehension (lower perplexity) with scalable runtime increase. Random insertion operation performs the best among all four approaches with a perplexity score of 5.} Therefore, we use the BERT model pre-trained with data from RI for further processes. This BERT model will be called Hardware Security-specific BERT (\textbf{HS-BERT}).

\subsection{Security Property Classification}
\label{sec:spg}

\begin{scriptsize}
\begin{table*}[!hb]
\renewcommand{\arraystretch}{1.5}
\caption{Sentence pre-processing methods and examples.}
\label{tab:preprocess}
\begin{adjustbox}{width=1\textwidth,center}
\begin{tabular}{cl}
\hline \hline
\multicolumn{1}{l}{}       & \textbf{Sentence}                                                                                                                        \\ \hline \hline
\textbf{Original}          & If KEYMGR is at the conclusion of the operation, KEYMGR.CTRL.STATUS stays in the same state, begins again.                               \\ \hline
\textbf{Verb Swapping}     & If KEYMGR is at the conclusion of the operation, KEYMGR.CTRL.STATUS \underline{[remain]} in the same state, \underline{[starts]} again.  \\ \hline
\textbf{Noun Swapping}     & If \underline{[condition]} is at the conclusion of the operation, \underline{[register]} stays in the same state, begins again.       \\ \hline
\textbf{Fragment Removal}  & If KEYMGR is at the conclusion of the operation, KEYMGR.CTRL.STATUS stays in the same state. \st{[ begins again.]}                       \\ \hline
\textbf{Fragment Addition} & If KEYMGR is at the conclusion of the operation, KEYMGR.CTRL.STATUS stays in the same state, \underline{[system sampling]} begins again. \\ \hline
\end{tabular}
\end{adjustbox}
\vspace{-0.1in}
\end{table*}
\end{scriptsize}

\begin{figure}[tb!]
\centering
  \includegraphics[width=0.95\linewidth]{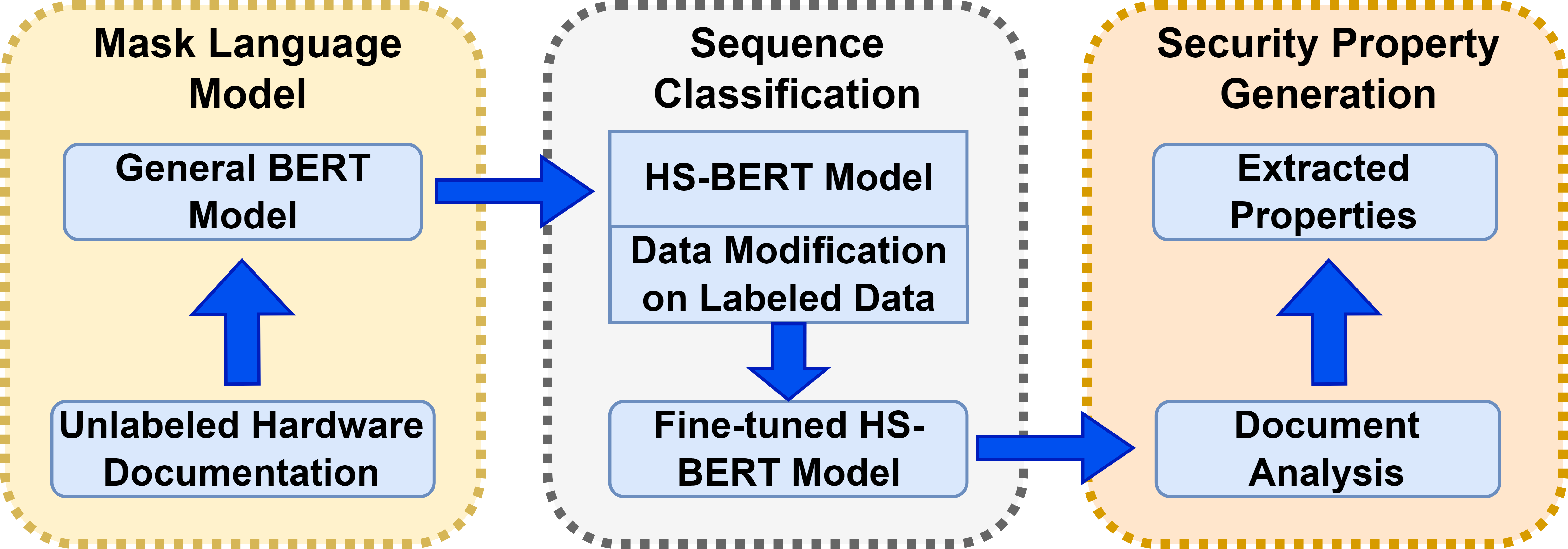}
  \caption{HS-BERT model training process.}
  \vspace{-0.1in}
  \label{fig:secgen}
\end{figure}

In this section, we will explain the process of security property classification from the design documentation utilizing the Sequence Classification Model (SCM), as mentioned in Section~\ref{sec:SCM}. 
\textcolor{black}{Figure~\ref{fig:secgen} illustrates the training procedure of NSPG framework, which consists of three stages. In the first stage, we obtain the HS-BERT model with sentences from hardware design documentation. In the second stage, we will modify the labeled ${\cal{D}}_{cls}$ with the HS-BERT model to fine-tune SCM. In the third stage, the HS-BERT model will be compared against other state-of-the-art BERT models.
Finally, we will validate each fine-tuned SCM with ${\cal{D}}_{val}$ and utilize the best-performing one to generate security properties from the documents. }

\textcolor{black}{${\cal{D}}_{cls}$ is used to fine-tune the SCM model and validate the performances of each classification model.
We will utilize ${\cal{D}}_{val}$ as unseen in-domain documentation, and evaluate the performances of the fine-tuned SCM model for security property generation}.
Each BERT sequence classification model is trained with different combinations of the original and modified training datasets to evaluate the classification performance based on accuracy, recall, precision, and F1-score.

Since our priority is to maximize the generation of security properties, we can tolerate the presence of a few non-property-related sentences (\emph{i.e.}, False Positives), which can be removed with further analysis. Therefore, accuracy and recall are the key metrics to determine the performance of sentence modification and SCM in NSPG. Higher values of accuracy and recall indicate better performance of the modification method and SCM model. The details will be discussed in Section~\ref{sec:trsec}.
In this work, we employ the SciBERT and general BERT model (``bert-base-uncased''), and compare their performances against HS-BERT in Section~\ref{sec:bertcom}.

\begin{figure*}[ht]
\centering
  \includegraphics[width=0.92\textwidth]{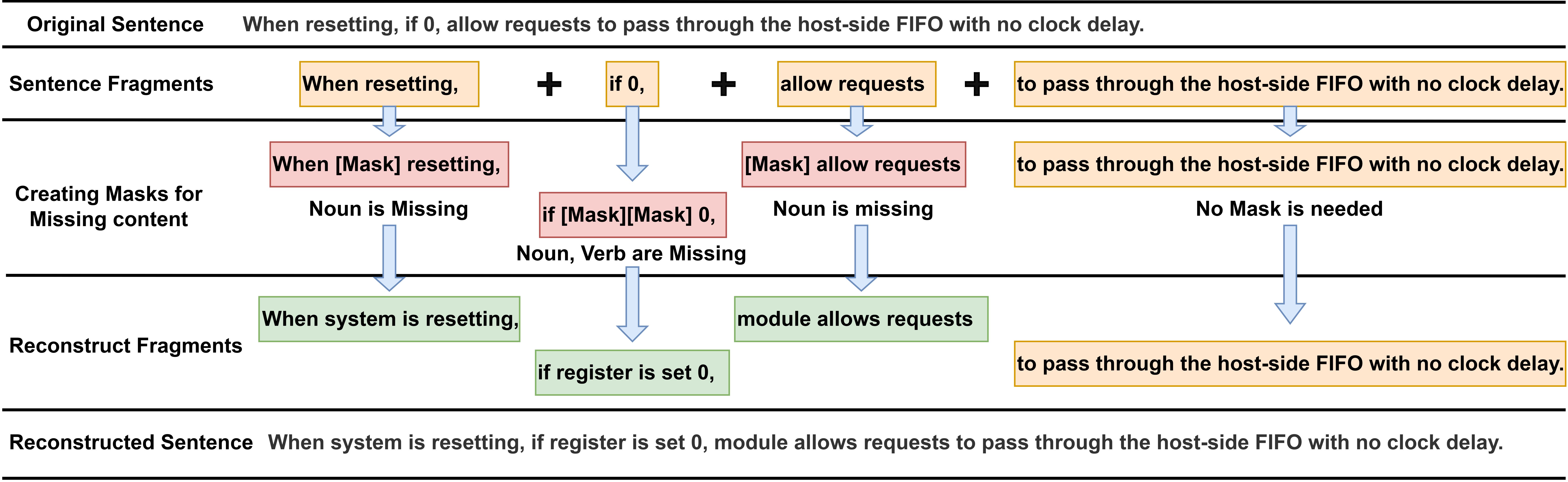}
  \caption{Sentence modification in documents. The first row represents the original sentence in the document. The second row shows each fragment of the original sentence. The third row adds [Mask] token into the fragment for missing components. The fourth row demonstrates the process of reconstructing the missing components of each fragment, and the last row shows the sentence after constructing the missing components.}
  \label{fig:reconstruct}
  \vspace{-0.1in}
\end{figure*}
\subsubsection{Data Modification for Property Classification}
\label{sec:datmod}
\begin{scriptsize}
\begin{table*}[!t]
\renewcommand{\arraystretch}{1.2}
\caption{Performance comparison between general BERT, SciBERT and HS-BERT with data modifications from ${\cal{D}}_{cls}$.}
\label{tab:BERTvs}
\begin{adjustbox}{width=0.95\textwidth,center}
\begin{tabular}{ccccccccccccc}
\hline \hline
\multicolumn{1}{l}{\textbf{}} & \multicolumn{4}{c}{\textbf{General BERT}}                                                                                                                    & \multicolumn{4}{c}{\textbf{SciBERT}}           & \multicolumn{4}{c}{\textbf{HS-BERT}}                                                                                                                    \\ \hline \hline
\multicolumn{1}{l}{\textbf{}} & \multicolumn{1}{l}{\textbf{Accuracy}} & \multicolumn{1}{l}{\textbf{Recall}} & \multicolumn{1}{l}{\textbf{Precision}} & \multicolumn{1}{l}{\textbf{F1-Score}} & \multicolumn{1}{l}{\textbf{Accuracy}} & \multicolumn{1}{l}{\textbf{Recall}} & \multicolumn{1}{l}{\textbf{Precision}} & \multicolumn{1}{l}{\textbf{F1-Score}}   & \multicolumn{1}{l}{\textbf{Accuracy}} & \multicolumn{1}{l}{\textbf{Recall}} & \multicolumn{1}{l}{\textbf{Precision}} & \multicolumn{1}{l}{\textbf{F1-Score}}  \\ \hline
\textbf{Baseline}                 & 83.1\%                                 & 81\%                                & 49.4\%                                  & 65.1\%                                 & 83.2\%                                 & 98\%                                & 46\%                                   & 61.7\%                     &  84\%        &   97\%     &    45\%       &  63\%                  \\ \hline
\textbf{MT}                   & 83.2\%                                 & 93.5\%                               & 51.8\%                                  & 65.8\%                                 & 82.2\%                                 & 96\%                                & 49\%                                   & 65\%                 &    85.1\%      &   97.3\%     &    48.7\%       &  64.7\%                       \\ \hline
\textbf{MOT}                  & 86\%                                  & 96.4\%                               & 46.7\%                                  & 63\%                                  & 88.1\%                                 & 98\%                                & 45\%                                   & 62\%                 & 90.1\%     & 98.3\%  & 49\%      & 65\%               \\ \hline
\textbf{MTT}                  & 81.2\%                                 & 83.4\%                               & 50.4\%                                  & 61.2\%                                 & 87.1\%                                 & 95\%                                & 44\%                                   & 60\%          &     88.1\%     &    97.5\%    &    48\%       &        64\%                 \\ \hline
\textbf{MOTMT}                 & 83.2\%                                 & 70.6\%                               & 35.5\%                                  & 71\%                                 & 87.5\%                                 & 97\%                                & 44\%                                   & 59\%            &    88.5\%      &    98.1\%    &     48\%      &      64\%                \\ \hline
\end{tabular}
\end{adjustbox}
\end{table*}
\end{scriptsize}
\textcolor{black}{Data pre-processing and data cleansing are utilized to generate unbiased data, which in turn is critical to furnish reliable machine learning algorithms~\cite{felix2019systematic}.} 
Studies have shown that biased learning can result from training on imbalanced or noisy data~\cite{stefanowski2016dealing, krawczyk2014cost}. Therefore, significant consideration should be provided to the data cleansing process and detailing the methods used in our studies. BERT model tokenizer is used to tokenize the sentences by its data pre-processing technique~\cite{devlin2018bert}.
In order to improve the performance of sequence classification, we propose a hardware domain-specific modification to further differentiate the features of each sentence in the documents, which enforces the BERT model to recognize more contextual dependencies. For instance, sentences including hardware operations and behavior are considered in-domain, while common descriptions are regarded as out-of-domain.

Although a plethora of data pre-processing approaches have been proposed to improve the performances of machine learning techniques in text mining and image classification~\cite{gupta2017set, iliou2015novel}, 
most language models are pre-trained with a wide range of documentation in a different domain.
In this work, we consider some data modification methods with hardware security-domain context and validate their impact on model performances. 
As shown in Table~\ref{tab:preprocess}, we apply several alternative methods, including verb swapping, noun swapping, fragment deletion, and fragment insertion, to preprocess all sentences in the dataset, described as follows: 

\noindent{\textbf{Verb Swapping}}: This approach substitutes the verbs in the sentence with semantic pre-defined hardware domain-related verbs. For example, in Table~\ref{tab:preprocess}, ``stay'' and ``begins'' in each sentence are replaced with ``remain'' and ``starts'' to formalize the usage of verbs.

\noindent{\textbf{Noun Swapping}}: This method intends to remove the arbitrary term used in each sentence, such as ``KEYMGR'' and ``KEYMGR.CTRL.STATUS'' in Table~\ref{tab:preprocess}, and replace them with more generalized terminology in the hardware domain such as ``condition'' and ``register''.

\noindent{\textbf{Fragment Deletion}}: \textcolor{black}{Some fragments of the sentence are not directly related to operation details. In this case, we intend to break the sentence into multiple conjunctions and remove the small fragments such as ``begin again'' in Table~\ref{tab:preprocess}, while retaining the more essential content of hardware operations.}

\noindent{\textbf{Fragment Insertion}}:  This approach adds information to the incomplete sentence. Typically, we consider a complete sentence consists of two (noun, verb) or three components (noun, verb, and noun). For each incomplete fragment in the sentence, we will construct them with a similar structure.

\subsubsection{Fragment Insertion Modification}
Through our initial analysis, the other data modification approaches furnished sub-par classification performance of up to 55\% accuracy, we have included and described the results pertaining to the fragment insertion technique.
Therefore, we determine that only fragment insertion improves the performance of the classification model. While noun swapping and verb swapping do not significantly affect performance, fragment removal degrades performance.
The primary reason is that fragment insertion adds more in-domain context to the original sentence, which helps the SCM identify the operation context.
Therefore, we can infer that instead of simplifying the data, the NLP model tends to require more information and complete structure in each sentence to learn the subjects more accurately. Intuitively, the prevailing assumption is that using more in-domain text in the training datasets should help with domain-specific classification.
Although each sentence in the document is constructed differently, we will explore this concept by adding domain-specific portions for each sentence. 

Figure~\ref{fig:reconstruct} shows an example of modifying a sentence in the document with fragment insertion. First, we break the sentence ``When resetting, if 0, allow requests to pass through the host-side FIFO with no clock delay.'' into fragments by identifying each conjunction ``When'', ``if'', and ``to'' in the sentence, and separating the sentence into fragments as: ``When resetting'', ``if 0'', ``allow requests'', and ``to pass through the host-side FIFO with no clock delay''. In these fragments, ``When resetting'', and ``allow requests'' are missing nouns, ``if 0'' is missing noun and verb, and ``to pass through the host-side FIFO with no clock delay'' is complete. We will add these missing components into the fragments by adding a [Mask] token into the fragment and applying the pre-trained HS-BERT to place appropriate in-domain terms for each [Mask] token. Hence,  ``When resetting'', ``if 0'', and ``allow requests'' will be transformed into ``When system is resetting'', ``if value is set 0'', and ``module allows requests''. 
\textcolor{black}{Since the other data modification approaches furnished sub-par classification performance of up to 55\% accuracy, we have included and described the results pertaining to the fragment insertion technique.}

\subsubsection{General BERT vs SciBERT vs HS-BERT}
\label{sec:bertcom}

\textcolor{black}{
In this section, we will discuss the available pre-trained BERT models and motivate our choice of the proposed HS-BERT, as shown in Section~\ref{sec:mlm}, for its inclusion in NSPG.
General BERT utilizes WordPiece~\cite{wu2016google} for unsupervised tokenization of input sequences, building its vocabulary with the most frequently used words or sub-word units. SciBERT, on the other hand, is constructed with a new WordPiece vocabulary on a scientific corpus using the SentencePiece1 library~\cite{sennrich-etal-2016-neural}. The token overlap between BERT and SciBERT vocabulary is 42\%, indicating a substantial difference in the frequently used words between scientific and general domain texts. {SciBERT was trained on a dataset comprising 1.14M papers from Semantic Scholar, where 82\% of the papers belong to the biomedical domain, while the remaining 18\% pertain to computer science~\cite{ammar2018construction}.}
}

Table~\ref{tab:BERTvs} shows the comparison between the general BERT model, SciBERT, and the proposed HS-BERT, training the same data modification method and the same 3927 samples from ${\cal{D}}_{cls}$. Another 500 samples from ${\cal{D}}_{cls}$ are used for validation. Modifications are applied to both in-domain and out-of-domain sentences. \ul{The baseline refers to no modification on both training and testing data.} \textcolor{black}{We have considered the following cases for modification:  \textbf{MT} refers to only modifying training data. \textbf{MTT} refers to modifying training and testing data. \ul{Note that the modification only changes the content of testing sentences and does not increase the size of test samples}. \textbf{MOT} refers to both modified and original training data. \textbf{MOTMT} refers to the original training data, its modified counterpart, and modified testing data.} \textcolor{black}{The results indicate that all three BERT models demonstrate improved performance in comparison to their baseline models, with an average increase of 0.6\% accuracy with General BERT, 3.1\% accuracy with SciBERT, and 3.95\% accuracy with HS-BERT.} For each data modification method, the largest effects of fine-tuning were observed in the MOT (+6.1\% accuracy with HS-BERT, +4.9\% accuracy with SciBERT, and +2.9\% accuracy with General BERT) and MOTOT (+4.9\% accuracy with SciBERT and +0.1\% accuracy with General BERT). While little effect was seen on recall, precision, and F1-score for HS-BERT and SciBERT, General BERT gains a significant improvement in recall when applying MOT (+15.4\%). However, HS-BERT, with fine-tuning, outperforms the state of art General BERT and SciBERT model on accuracy and recall, while performing similarly on precision and F1-score. \textcolor{black}{Based on these observations, we determine that HS-BERT is more suitable for extracting potential security properties, which we have subsequently incorporated in NSPG.}

\begin{scriptsize}
\begin{table}[t]
\renewcommand{\arraystretch}{1.3}
\caption{Sequence Classification for ${\cal{D}}_{val}$. Each row represents the accuracy, recall, and F1-score for different HS-BERT models and different modified labeled datasets.}
\label{tab:SC}
\begin{adjustbox}{width=1\columnwidth,center}
\begin{tabular}{ccccccc}
\hline \hline
\multicolumn{1}{l}{}   & \multicolumn{2}{c}{\textbf{OpenTitan}}                    & \multicolumn{2}{c}{\textbf{RISCV}}                        & \multicolumn{2}{c}{\textbf{Openrisc}}                     \\ \hline \hline
\multicolumn{1}{l}{}   & \multicolumn{1}{l}{Accuracy} & \multicolumn{1}{l}{Recall} & \multicolumn{1}{l}{Accuracy} & \multicolumn{1}{l}{Recall} & \multicolumn{1}{l}{Accuracy} & \multicolumn{1}{l}{Recall} \\ \hline
\textbf{Base-HS-BERT}  & 70.6\%                       & 72.8\%                     & 71.9\%                       & 71.2\%                     & 80.3\%                       & 80.2\%                     \\ \hline
\textbf{MT-HS-BERT}    & 74.5\%                       & 80.4\%                     & 75.8\%                       & 79.2\%                     & 83.6\%                       & 82.6\%                     \\ \hline
\textbf{MOT-HS-BERT}   & 81.5\%                       & 93\%                       & 79.1\%                       & 90.1\%                     & 88.3\%                       & 87\%                       \\ \hline
\textbf{MTT-HS-BERT}   & 75\%                         & 85\%                       & 74.3\%                       & 83.5\%                     & 86.3\%                       & 82.6\%                     \\ \hline
\textbf{MOTMT-HS-BERT} & 76\%                         & 84.1\%                     & 73.3\%                       & 80.5\%                     & 87\%                         & 84.7\%                     \\ \hline
\end{tabular}
\end{adjustbox}
\end{table}
\end{scriptsize}

\subsubsection{Fine-tuning SCM model}
\label{sec:trsec}

In this stage, NSPG will apply fine-tuned HS-BERT SCM to identify the security properties in the SoC documentation.
In order to emulate this scenario in which the SCM is applied on an unseen document to determine whether the sentence is a security property or not, we will test each trained sequence classification model on ${\cal{D}}_{val}$.
Table~\ref{tab:SC} shows the results for HS-BERT performance with no modification and the data modification approach, as described in Section~\ref{sec:bertcom}. \ul{Base-HS-BERT refers to the SCM performance fine-tuned with no data modification.}
The HS-BERT model trained with the MOT modification performs the best with an average of 82\% accuracy, and 90\% recall score. \textcolor{black}{Since this method consists of both the original and the modified sentences, it improves the diversity of the training dataset, which helps the model to learn more features and context of a property-related sentence. Therefore, we will use MOT modification as our final data modification approach for the SCM model.} 


\setlength{\textfloatsep}{3pt}
\begin{algorithm}[t!]
\vspace{0.7mm}
\raggedright{\textbf{Input}: Document Files, Labeled Dataset\\
\textbf{Output}: \textcolor{black}{Property, Non-Property}}
 \caption{Sentence Formalization}
\vspace{1mm}
\begin{algorithmic}[1]
\State \textbf{MLM}.train(Document Files)
\State \textbf{Initialize} Enhanced Labeled Dataset
\For {each \textbf{Sentence} in \textbf{Labeled Dataset}}
\State   Split(\textbf{Sentence}) $\rightarrow$ \textbf{Fragments}
\For{each \textbf{Fragment} in \textbf{Fragments}}
\If {\textbf{Fragment} is \textbf{incomplete}}
\State Apply \textbf{Data Modification} $\rightarrow$ \textbf{Fragment}
\EndIf
\EndFor
\State Join(\textbf{Fragments}) $\rightarrow$ \textbf{Sentence}
\State \textbf{Append} Enhanced Labeled Dataset$\leftarrow$\textbf{Sentence}
\EndFor
\State \textbf{SCM}.train(Enhanced Labeled Dataset)
\For {each \textbf{Sentence} in \textbf{Document}}
\State \textbf{if} SC.predict(\textbf{Sentence}) $>$ 0.5 \textbf{then}
\State \ \ \ \ \textbf{Sentence} $\rightarrow$ \textbf{Property}
\State \textbf{else}
\State \ \ \ \ \textbf{Sentence} $\rightarrow$ \textbf{Non-Property}
\EndFor
\end{algorithmic}
\label{algo:1}
\end{algorithm}


\subsubsection{NSPG Framework Summary}

After validating each procedure in training, we will construct the pipeline of our proposed framework NSPG. It comprises data augmentation, data modification, fine-tuning masked language model, and sequence classification model. The fine-tuned sequence classification model is used to analyze the unseen documents, and extract the security properties which contain essential information about operation behaviors, and register interactions. Algorithm~\ref{algo:1} demonstrates the details of each stage for the training and security property generation process. These extracted security properties can be further utilized by commercial hardware verification approaches, such as Cadence Jaspergold, and detect potential vulnerabilities in the design. 
The generated security properties as well as their application in detecting SoC vulnerabilities will be evaluated in Section~\ref{experiment}.

\section{OpenTitan SoC and Threat Model}
\label{threat}

\subsection{OpenTitan SoC}

Since we do not have access to commercial designs, our framework has been developed using open-source SoCs, such as OpenSPARC, MIPS, OpenRISC, RISC-V and OpenTitan.
In this work, while the OpenTitan design documentation is utilized for both training and validation, the other documents are only used for training. The OpenTitan document considered in our study corresponds to a buggy design, which was used in the Hack@DAC 2022 hardware hacking competition~\cite{HACKDAC211:online}. 

\subsection{Threat Model}
\label{sec:threat}

Our threat model is similar to the one used in Hack@DAC 2022. The attack scenarios considered are as follows:
 \begin{itemize}
     \item \textcolor{black}{An unprivileged software adversary that can access the core in user mode, with control over user-space interfaces and the ability to issue unprivileged system instructions and request feedback.}
    \item A privileged software adversary that can execute malicious code with supervisor privilege but may target higher privilege levels or bypass security countermeasures.
    \item An authorized debug adversary which is capable of unlocking and debugging production devices.
 \end{itemize}
The aim of the adversary is to exploit any potential vulnerabilities in the SoC in order to bypass any security objective.

\subsection{Security Objectives and Features}
\label{sec:objective}

The security objectives of OpenTitan are as follows~\cite{HACKDAC211:online}:
\begin{itemize}
    \item \textit{\textbf{SO1}}: \textcolor{black}{Preventing privilege level compromise due to unprivileged code running in the core.}
    \item \textbf{\textit{SO2}}: Protecting system debug interface from malicious or unauthorized debugger.
    \item \textit{\textbf{SO3}}: Protecting device integrity and preventing exploits from a software adversary.
\end{itemize}
Various security features have been developed for the OpenTitan SoC to support the aforementioned security objectives, including privileged access control to peripherals, write and read locks on registers to protect the privilege integrity from unprivileged instruction, and reset to flush sensitive information in cryptographic processors. 
\section{Experiments}
\label{experiment}

In this section, we will demonstrate the evaluation of our framework NSPG with five OpenTitan IP documents, and discuss its performance on security property extraction. Furthermore, these extracted properties will be utilized with verification methods to detect vulnerabilities in the design.

\subsection{Experimental Setup}
\label{sec:setup}
\begin{scriptsize}
\begin{table}[t!]
\caption{Number of processed sentences, security properties generated, and properties covered and not covered by design verification (DV) documentation.}
\label{tab:genlist}
\begin{adjustbox}{width=1\columnwidth,center}
\begin{tabular}{cccccc}
\hline \hline
\textbf{Hardware IP}    & \textbf{Sentences} & \textbf{Extracted} & \textbf{Properties} & \textbf{\begin{tabular}[c]{@{}c@{}}Covered\\ by DV\end{tabular}} & \textbf{\begin{tabular}[c]{@{}c@{}}Not covered\\  by DV\end{tabular}} \\ \hline \hline
\textbf{Key Manager}    & 241                & 51                 & 47                  & 40                                                               & 7                                                                     \\ \hline
\textbf{LC Controller}  & 375                & 79                 & 76                  & 65                                                               & 11                                                                    \\ \hline
\textbf{HMAC}           & 170                & 28                 & 27                  & 19                                                               & 8                                                                     \\ \hline
\textbf{KMAC}           & 367                & 84                 & 74                  & 60                                                               & 14                                                                    \\ \hline
\textbf{OTP Controller} & 570                & 105                & 102                 & 84                                                               & 18                                                                    \\ \hline
\textbf{Total}          & 1723               & 347                & 326                 & 268                                                              & 58                                                                    \\ \hline
\end{tabular}
\end{adjustbox}
\end{table}
\end{scriptsize}
\begin{scriptsize}
\begin{table*}[t]
\renewcommand{\arraystretch}{1.2}
\caption{Examples of security properties generated from the framework.}
\label{tab:buglist}
\begin{adjustbox}{width=0.95\textwidth,center}
\begin{tabular}{clc}
\hline \hline
\textbf{Module}         & \textbf{Constructed Security Properties}                                                                                                                                                                                           & \textbf{HW CWE Category}                                                                                             \\ \hline \hline
\textbf{Key Manager}    & \begin{tabular}[c]{@{}l@{}}Upon Disabled entry, the internal key is updated with KMAC computed random values; however, \\ previously generated sideload key slots and software key slots are preserved.\end{tabular}               & Improper Zeroization of Hardware Register - (1239)                                                                   \\ \cline{2-3} 
\textbf{}               & \begin{tabular}[c]{@{}l@{}}Invalid state is entered whenever key manager is deactivated through the life cycle connection or \\ when an operation encounters a fault .\end{tabular}                                                & \begin{tabular}[c]{@{}c@{}}Improper Finite State Machines (FSMs) in \\ Hardware Logic - (1245)\end{tabular}          \\ \cline{2-3} 
\textbf{}               & When an illegal operation is supplied, the err\_code is updated and the operation is flagged as done with error.                                                                                                                   & \begin{tabular}[c]{@{}c@{}}Improper Protection for Outbound Error Messages\\ and Alert Signals - (1320)\end{tabular} \\ \cline{2-3} 
\textbf{}               & When the life cycle controller deactivates the key manager, the key manager transitions to the Invalid state.                                                                                                                      & \begin{tabular}[c]{@{}c@{}}Improper Finite State Machines (FSMs) in \\ Hardware Logic - (1245)\end{tabular}          \\ \hline
\textbf{LC Controller}  & fatal\_bus\_integ\_error\_q is triggered when a fatal TL-UL bus integrity fault is detected.                                                                                                                                       & \begin{tabular}[c]{@{}c@{}}Improper Protection for Outbound Error Messages\\ and Alert Signals - (1320)\end{tabular} \\ \cline{2-3} 
\textbf{}               & fatal\_bus\_integ\_error\_q is set to 1 if a fatal bus integrity fault is detected.                                                                                                                                                & \begin{tabular}[c]{@{}c@{}}Improper Protection for Outbound Error Messages\\ and Alert Signals - (1320)\end{tabular} \\ \hline
\textbf{HMAC}           & If SW wants to convert the message byte order, SW should set CFG.endian\_swap to 1.                                                                                                                                                & Expected Behavior Violation - (440)                                                                                  \\ \cline{2-3} 
\textbf{}               & When the SHA engine is disabled the digest is cleared.                                                                                                                                                                             & \begin{tabular}[c]{@{}c@{}}Sensitive Information in Resource Not Removed\\ Before Reuse - (226)\end{tabular}         \\ \cline{2-3} 
\textbf{}               & \begin{tabular}[c]{@{}l@{}}If CPU writes value into the register, the value is used to randomize internal variables such as secret\\  key, internal state machine, or hash value.\end{tabular}                                     & \begin{tabular}[c]{@{}c@{}}Sensitive Information in Resource Not Removed\\ Before Reuse - (226)\end{tabular}         \\ \hline
\textbf{KMAC}           & If the EnMasking parameter is not set, the second share is always zero.                                                                                                                                                            & Improper Zeroization of Hardware Register - (1239)                                                                   \\ \cline{2-3} 
\textbf{}               & If EnMasking is not defined, the KMAC merges the shared key to the unmasked form before uses the key.                                                                                                                              & \begin{tabular}[c]{@{}c@{}}Sensitive Information in Resource Not Removed\\ Before Reuse - (226)\end{tabular}         \\ \cline{2-3} 
\textbf{}               & \begin{tabular}[c]{@{}l@{}}If the EnMasking parameter is set and CFG\_SHADOWED.msg\_mask is enabled, the message is masked \\ upon loading into the Keccak core using the internal entropy generator.\end{tabular}                   & Security Primitives and Cryptography Issues - (1205)                                                                 \\ \hline
\textbf{OTP Controller} & \begin{tabular}[c]{@{}l@{}}The otp\_lc\_data\_o.secrets\_valid signal is a multibit valid signal that is set to lc\_ctrl\_pkg::On if the \\ SECRET2 partition containing the root keys has been locked with a digest.\end{tabular} & Improper Prevention of Lock Bit Modification - (1231)                                                                \\ \cline{2-3} 
\textbf{}               & Read transactions through the CSR window will error out if they are out of bounds, or if read access is locked.                                                                                                                    & Improper Access Control for Register Interface - (1262)                                                              \\ \hline
\end{tabular}
\end{adjustbox}
\vspace{-0.1in}
\end{table*}
\end{scriptsize}
All of our experiments are run on a server consisting of 40 CPUs of 64-bit Intel(R) Xeon(R) E5-2698 v4 @ 2.20GHz. Our NSPG framework is implemented in Python. We intend to release our framework in GitHub soon, for use by other researchers. 
The entire SoC documentation comprises of 33 IP design specifications, \textcolor{black}{with 10865 sentences.} Each IP design documentation demonstrates information on register descriptions, functionalities, and operation processes, including the security features for various modules, that will be used for evaluating NSPG~\cite{OpenTita77:online}. 
\textcolor{black}{Among these, the contents of the operational processes or behaviors can be transformed into security properties, while the others are treated as non-properties.} 
\textcolor{black}{In the following subsections, we will evaluate each extracted sentence from five unseen IP documents and identify whether they can be transformed into security properties.} Moreover, we will utilize these security properties to search for potential violations in relevant hardware IPs. \textcolor{black}{Since the list of registers in SoC design is available to us, we craft the detailed security properties from the IP security specification and checklist of the OpenTitan SoC registers.}
\textcolor{black}{The verification of SoC is based on the threat model and security objectives, as described in Section~\ref{sec:threat} and~\ref{sec:objective}, respectively.  }

\subsection{NSPG Framework Evaluation}
\label{exp:eval}


First, NSPG processes each text file consisting of sentences in the design documentation. It filters out any sentence having less than 10 words, since they usually do not contain enough information about operation behaviors. The rest of the sentences will be parsed through the trained HS-BERT sequence classification model discussed in Section~\ref{method}, and the extracted sentences from each IP document will be listed in property text files. 
As shown in Table~\ref{tab:genlist}, 1723 sentences are processed, and 344 sentences are extracted as potential security properties.
Overall, 326 sentences (94\% of the 347 extracted sentences) can be utilized to generate security properties for design validation. \textcolor{black}{We compare the generated security properties with the test cases listed in the Design Verification (DV) documentation, which describes all the test cases and IP operations needed to be checked by the SoC designer. \textbf{While 268 of our generated security properties are covered in DV, 58 properties are not covered in the test cases, which clearly demonstrate that NSPG is adept at accounting for specifications that are not covered in DV.}}
This shows that our proposed framework, NSPG, is able to efficiently identify security properties in the documents. Next, we use these newly generated security properties to verify the bugs.

\subsection{Effectiveness in Discovering Violations}
\label{exp:search}

\begin{figure}[tb!]
\centering
  \includegraphics[width=1\linewidth]{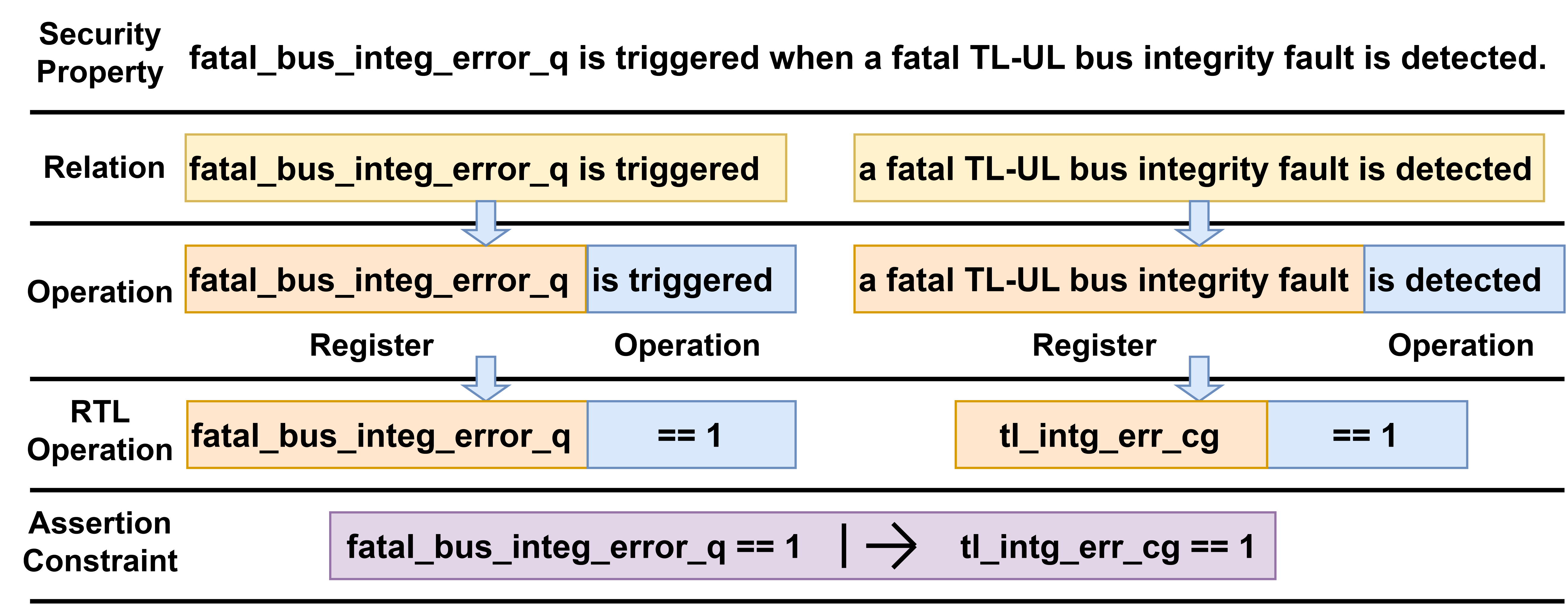}
  \caption{Process of transforming a security property into an assertion. The first row breaks the sentences into relation conjunctions. The second row identifies the design components and the operations. The third row converts the texts into registers and operators. The forth row reconstructs them into the constraint for verification.}
  \label{fig:trs}
\end{figure}

\begin{scriptsize}
\begin{table*}[t!]
\renewcommand{\arraystretch}{1.5}
\caption{Eight bugs found in Key Manager, LC controller, HMAC, KMAC, and OTP memory controller, the violated security objectives, CWE, CVSS, and security impacts.}
\label{tab:allbugs}
\begin{adjustbox}{width=1\textwidth,center}
\begin{tabular}{lccccc}
\hline \hline
\textbf{Vulnerability No.}                                      & \textbf{Module}       & \textbf{Sec Obj Violated} & \textbf{CWE Category}                                                    & \textbf{CVSS}~\cite{NVDCVSSv94:online} & \textbf{Security Impact}      \\ \hline \hline
Bug 1 - Secret key is not wiped under a invalid state.          & Key Manager           & Device Integrity            &   Sensitive Information in Resource Not Removed Before Reuse - (226)          & 3.3            & Information Leakage  \\ \hline
Bug 2 - Secret key is not wiped during the operation state.     & Key Manager           & Device Integrity            &  Sensitive Information Not Removed Before Reuse - (226)          & 4.4            & Information Leakage  \\ \hline
Bug 3 - JTAG does not support bus integrity checks.             & LC Controller         & Device Integrity            & Improper Protection for Outbound Error Messages and Alert Signals - (1320)  & 4              & Unexpected Behavior  \\ \hline
Bug 4 - The message byte order conversion is not operational.   & HMAC                  & Device Integrity            & Expected Behavior Violation - (440)    & 3.1            & Unexpected Behavior  \\ \hline
Bug 5 - Digest is not cleared when SHA is disabled.             & HMAC                  & Device Integrity            & Sensitive Information in Resource Not Removed Before Reuse - (226)                            & 4.8            & Information Leakage  \\ \hline
Bug 6 - Key is not written with randomly generated value.        & HMAC                  & Exploits from Software      &    Sensitive Information in Resource Not Removed Before Reuse - (226)                        & 2.8            & Unprivileged Access \\ \hline
Bug 7 - Software does not provide the key in masked form.       & KMAC                  & Exploits from Software      & Security Primitives and Cryptography Issues - (1205)                                   & 4.2            & Information Leakage  \\ \hline
Bug 8 - Lock control signal is bypassed with fault instruction. & OTP Controller & Unprivileged Code           & Improper Access Control for Register Interface - (1262)                     & 5.8            & Unprivileged Access \\ \hline
\end{tabular}
\end{adjustbox}
\vspace{-0.1in}
\end{table*}
\end{scriptsize}
\textcolor{black}{The extracted security properties provide us with essential information to construct various constraints when generating test cases for vulnerable IP designs. We choose to transfer the properties into SystemVerilog assertion format.
Figure~\ref{fig:trs} shows an example of the process, transforming an extracted property into an assertion that can be applied for verification. It breaks the conjunctions of the sentence to identify the fragments of the operation relation. The nouns are transformed into RTL-level registers based on the IP register listing, and the verbs are converted into operators. Finally, the fragments rejoin to present a constraint for operation behavior, which can be asserted into the original RTL.
Table~\ref{tab:buglist} outlines a few security properties generated by our framework and their corresponding CWE categories, from which we discover some vulnerabilities in the design. 
By creating constraints based on the acquired security properties and generating test cases on the design, we are able to detect eight vulnerabilities in the buggy IP designs.
\textcolor{black}{Table~\ref{tab:allbugs} demonstrates the details of eight bugs we identified from the extracted security properties. It presents the security objectives violated, CVSS score~\cite{NVDCVSSv94:online} (ranges from 0 to 10, a higher score refers to more severe vulnerabilities), CWE categories, and the potential security impacts on the system. We will discuss these vulnerabilities and their impacts on the system as follows.}
}
\smallskip

\noindent
\textbf{A. Key Manager:}
{The key manager implements the root key operation for the system and allows it to protect critical assets from malicious software.} Two vulnerabilities are found in this IP: (1) The security property requires the key manager to wipe internal storage when it is in an invalid state. However, the implementation reverses the operation and wipes the key under a valid state. (2) It is required to continuously wipe the secret key with entropy during the operation state. However, the implementation does not replace the key registers, leaving sensitive information vulnerable. These vulnerabilities could potentially impact the confidentiality of the secret key, allowing the attacker to reveal the key information.

\smallskip

\noindent
\textbf{B. LC Controller:}
The life cycle (LC) controller controls the peripheral interactions on the chip interconnect bus. 
The security property requires the signal \textit{fatal\_bus\_integ\_error\_q} to be set to one, when any bus integrity fault is detected. However, the boundary-scan test controller (JTAG) Test Access Port (TAP) does not provide a bus integrity check signal, which might cause integrity check failure in the life cycle controller and unexpected behavior in the system.

\smallskip

\noindent
\textbf{C. HMAC:}
\textcolor{black}{The HMAC module implements a SHA-256 hash-based authentication module to ensure the integrity of any incoming message and its encryption code from the secret key.} Three security bugs are found in this IP: (1) The first bug occurs when the software wants to convert the message byte order. It is required to set \textit{CFG.endian\_swap} register to one. However, the implementation reverses the operation, making the IP convert the message, when \textit{CFG.endian\_swap} is zero. This will cause unexpected operations due to incorrect instructions for converting messages. (2) The second bug occurs when the SHA engine is disabled. Although it is required to clear the digest in HMAC, no implementation is built to satisfy this property. (3) The third bug involves key randomization when the CPU writes values to the secret key. The key is required to be wiped with randomly generated value; however, no implementation is addressed for this property. 

\smallskip

\noindent
\textbf{D. KMAC:} 
\textcolor{black}{The KMAC module is a Keccak-based message authentication code generator used to verify the incoming messages. It utilizes masked permutations to prevent side-channel attacks.} The security property requires the software to provide the key in masked form when the \textit{EnMasking} parameter is not set and the \textit{SwKeyMasked} parameter is set. However, this mechanism is not correctly implemented, leaving the software with an unmasked key. This could cause key leakage through an unprivileged software adversary. 

\smallskip

\noindent
\textbf{E. OTP Memory Controller:}
The OTP memory controller is a module that provides a device with a one-time programming functionality. The security property only allows the IP to respond and write into the readout register when the lock control is not active. However, the IP is implemented with a mechanism that allows it to bypass the read-and-write lock control signal every four clock cycles. This will cause unpredictable behavior of the controller and allow the software adversary to attack the integrity of the module.

\smallskip

In summary, we have discovered eight vulnerabilities in five hardware IP designs \textcolor{black}{from the 326 security properties we generated using NSPG.}
These vulnerabilities may cause information leakage, unexpected behavior, and unprivileged accesses in these IPs. \textcolor{black}{It proves that these extracted security properties can provide valuable information to generate constraints for the hardware verification process. }


\section{Discussion}
\label{discussion}
\noindent
\textbf{A. False Positives and False Negatives:}  We will show some examples of the false positives and false negatives that were encountered, represented as FP and FN, respectively. A few of the FPs are short sentences that furnish equations or incomplete register interactions such as  ``How often FSM wakes up from ADC PD mode to take a sample, measured in always on clock cycles.''
Although the framework is able to identify the operations, there is not enough information in the sentence to construct a complete security property.
In contrast, FNs usually involve the structure of the sentences such as ``To this end, the processor has to set the SIDELOAD bit in CTRL\_SHADOWED to 1''.  Despite including two registers, the operation does not describe their behaviors separately, making it difficult for NSPG to classify them.
\textcolor{black}{Overall, the FP rate is  4\%, and FN rate is 15\% for the five IP documents.} It should be mentioned that NSPG is the first work that utilizes NLP for hardware security property generation, and there are potentials for further improving its performance.

\smallskip
\noindent
\textbf{B. Compared to text classification models:} {We also explore standard text classification models such as TF-IDF and Bag-of-Words trained with ${\cal{D}}_{cls}$ and tested on ${\cal{D}}_{val}$. Among them, TF-IDF achieved 70\% accuracy for OpenTitan, and 61\% accuracy for RISC-V, Bag-of-Words achieved 51\% accuracy for OpenTitan, and 53.12\% for RISC-V.} Hence, we can conclude that HS-BERT is significantly more effective than standard text classifiers.

\smallskip

\noindent
\textbf{C. Compared to ChatGPT:} \textcolor{black}{
\textcolor{black}{In this section, we compare NSPG against ChatGPT~\cite{Introduc18:online}, a popular chatbot based on the OpenAI Generative Pre-trained Transformer (GPT)-3.5 language model that is adept at generating human-like text responses to any input in a conversational context~\cite{schulman2022chatgpt}.} \textcolor{black}{
We investigated if such a state-of-the-art off-the-shelf general LLM can outperform a small domain-specific BERT model in solving the security property identification task. We evaluated their performances on a reduced dataset of 50 sentences, which contained 25 property-related and 25 non-property-related sentences (the dataset size was necessitated by resource limitations, since GPT-3.5 is not publicly available as an open-source model). ChatGPT's evaluation resulted in numerous false positives and false negatives, with only 35 sentences correctly classified, many of which were unsuitable for SoC security verification.} For example, sentences like ``The ADC is continually powered on" and ``In addition, they could potentially also be extracted when being transferred over the TL-UL bus interface" were incorrectly classified as properties. Conversely, sentences such as ``For encryption or if the mode is set to CFB, OFB, or CTR, there is no such initial delay upon changing the key" and ``The AES unit cannot recover from such an error and needs to be reset" were labeled as non-properties. The accuracy, recall and F1-score obtained by the ChatGPT model were 68\%, 88\% and 73\%, respectively. \textcolor{black}{On the contrary, NSPG outperforms ChatGPT by identifying all 25 property-related sentences, thereby furnishing an accuracy, recall, and F-1 score of 100\%, respectively.}
}

\section{Related Work}
\label{related}

This section presents prior related work proposed for hardware security property generation and verification.
\subsection{Hardware Security Verification Approaches}
\label{sec:sec_app}
\textcolor{black}{\textit{Information Flow Tracking} (IFT) improves hardware security by identifying malicious input interfaces, regulating the use of counterfeit information, and verifying hardware operation flows~\cite{suh2004secure, tiwari2009complete, oberg2010theoretical, ardeshiricham2017register, qin2020formal, ardeshiricham2019verisketch}.} This approach has been demonstrated to effectively detect malicious attacks with security property-generated constraints.
\textit{Theorem proving} utilizes a language, VeriCoq, to automate the transformation of RTL into Coq theorem. This approach validates the converted language through the Proof-Carrying Hardware Intellectual Property (PCHIP) framework and ensures the robustness of a third-party IP with formal proofs of security properties~\cite{love2011proof,bidmeshki2015vericoq}.
\textit{Assertion-based security verification} employs formal and simulation-based methods to detect violations against assertions included in RTL designs~\cite{bilzor2012evaluating, hicks2015specs}. 
\textit{Property-specific information flow} utilizes property specifications to generate information flow models that are catered to the pre-defined security properties. The models are verified by conducting a property-specific search and identifying security-critical paths~\cite{hu2018property}. 
\textit{Directed test generation} has been developed to address state space explosion. Existing research, such as~\cite{ahmed2018directed, lyu2020automated}, utilize control flow graphs on RTL models to ensure functional correctness. Symbolic techniques incorporating security properties and concolic testing can detect violations in RTL and conclusively verify designs~\cite{meng2021rtl, zhang2018end}. 


\subsection{Property Generation Techniques}

A recent approach, \textit{Isadora}, creates information flow specifications of core designs and combines IFT and security specifications~\cite{deutschbein2021isadora}. 
It requires conclusive testbenches and simulation traces for property generation. Therefore, it might be not scalable when applied to more complex SoCs.
PCHIP introduces the concept of theorem generation functions, which enables the generation of security theorems independent from the information flow traces, thereby assisting in the development of data secrecy properties~\cite{jin2017data, bidmeshki2017data}. 
However, PCHIP is only limited to cryptographic circuits.
\textcolor{black}{\textit{SCIFinder} is a recent approach that generates Security-Critical Invariants (SCI) for design verification~\cite{zhang2017identifying}.} 
This approach only studies a limited size of security properties and is not scalable when applied to property generation on more complex processors.
\textit{Comments-based property generation} is an NLP-based translation technique was developed, which automatically generates Computation Tree Logic (CTL) verification properties from Hardware Description Language (HDL) \textit{code comments}~\cite{harris2015generating}. 
In comparison, NSPG uses a BERT model to automatically identify and generate new security property-related sentences from the documentation. Furthermore, \cite{harris2015generating} has been evaluated only on small \emph{well-commented} benchmarks. 


\section{Conclusion}
\label{conculsion}

\balance

\textcolor{black}{This paper presents NSPG, the first NLP-based automated hardware security property generation method, that utilizes SoC documentation to extract security properties. NSPG includes a novel hardware security-specific language model (HS-BERT) and a data modification technique to improve automated security property generation.} Our framework is evaluated on OpenTitan SoC documentation, \textcolor{black}{resulting in 326 correctly extracted security properties from 1723 sentences for five hardware IPs.}
Furthermore, these security properties help discover eight vulnerabilities in a buggy design, which proves the effectiveness of the generated security properties.
Additionally, our evaluations prove that NSPG furnishes better performance than ChatGPT, a popular chatbot system, for SoC security property generation. With the advent of LLMs, we envision that NSPG will lay the foundation for utilizing NLP approaches in SoC design and verification. 

\section{Acknowledgement}
\label{Acknowledgement}

\balance

This research is partially supported by Technology Innovation Institute, Abu Dhabi,  
United Arab Emirates.



\bibliographystyle{IEEEtranS}
\bibliography{reference}

\end{document}